\documentclass[runningheads]{llncs}
\usepackage{url}
\usepackage[dvipsnames]{xcolor}
\usepackage{graphicx}
\begin{document}
\title{ImageCHD: A 3D Computed Tomography Image Dataset for Classification of Congenital Heart Disease}
%\author{For Blind Review}
%
%\authorrunning{For Blind Review}
\author{
Xiaowei Xu \inst{1}  \and 
Tianchen Wang \inst{2} \and 
Jian Zhuang \inst{1} \and
Haiyun Yuan\inst{1} \and 
Meiping Huang\inst{1} \and 
Jianzheng Cen \inst{1} \and
Qianjun Jia \inst{1} \and
Yuhao Dong \inst{1} \and
Yiyu Shi\inst{2} 
}
% index{Xu, Xiaowei}
% index{Wang, Tianchen}
% index{Jian Zhuang}
% index{Haiyun Yuan}
% index{Meiping Huang}
% index{Jianzheng Cen}
% index{Qianjun Jia}
% index{Yuhao Dong}
% index{Shi, Yiyu}
\authorrunning{X. Xu, et al.}
\institute{
Guangdong Provincial People's Hospital, \\
\email{xiao.wei.xu@foxmail.com, huangmeiping@126.com, zhuangjian5413@tom.com}
\and University of Notre Dame\\
\email{{\{twang9, yshi4\}@nd.edu}}
}
\maketitle              % typeset the header of the contribution
\begin{abstract}
Congenital heart disease (CHD) is the most common type of birth defects, which occurs 1 in every 110 births in the United States.
CHD usually comes with severe variations in heart structure and great artery connections that can be classified into many types. Thus 
highly specialized domain knowledge and time-consuming human process is 
needed to analyze the associated medical images. 
On the other hand,
due to the complexity of CHD and the lack of dataset, little has been explored on the automatic diagnosis (classification) of CHDs. 
In this paper, we present ImageCHD, the first medical image dataset for CHD classification. 
ImageCHD contains 110 3D Computed Tomography (CT) images covering 
most types of CHD, which is of decent size 
compared with existing medical imaging datasets. 
%Each image is labelled by a team of four experienced cardiovascular radiologists.
Classification of CHDs requires the identification of 
large structural changes without any local tissue changes, 
with limited data. It is an example of a larger class of problems
that are quite difficult for current machine-learning based 
vision methods to solve. 
To demonstrate this, 
we further present a baseline framework 
for automatic classification of CHD, 
based on a state-of-the-art CHD segmentation method. 
Experimental results show that the baseline framework 
can only achieve a classification accuracy of 81.9\% under selective prediction scheme with 88.8\% coverage,
leaving big room for further improvement.
We hope that ImageCHD can stimulate further 
research and lead to innovative and generic solutions that 
would have an impact in multiple domains.
Our dataset is released to the public \cite{ourdataset}.

\end{abstract}
\section{Introduction}
Congenital heart disease (CHD) is the problem with the heart structure that is present at birth, which is the most common type of birth defects \cite{bhat2016illustrated}.
In recent years, noninvasive imaging techniques such as computed tomography (CT) have prevailed in comprehensive diagnosis, intervention decision-making, and regular follow-up for CHD.
However, analysis (e.g., segmentation or classification) of these medical images are usually performed manually by experienced cardiovascular radiologists, which is time-consuming and requires highly specialized 
domain knowledge.

\begin{figure}[!htb]
\centering
\includegraphics[width=0.95\textwidth]{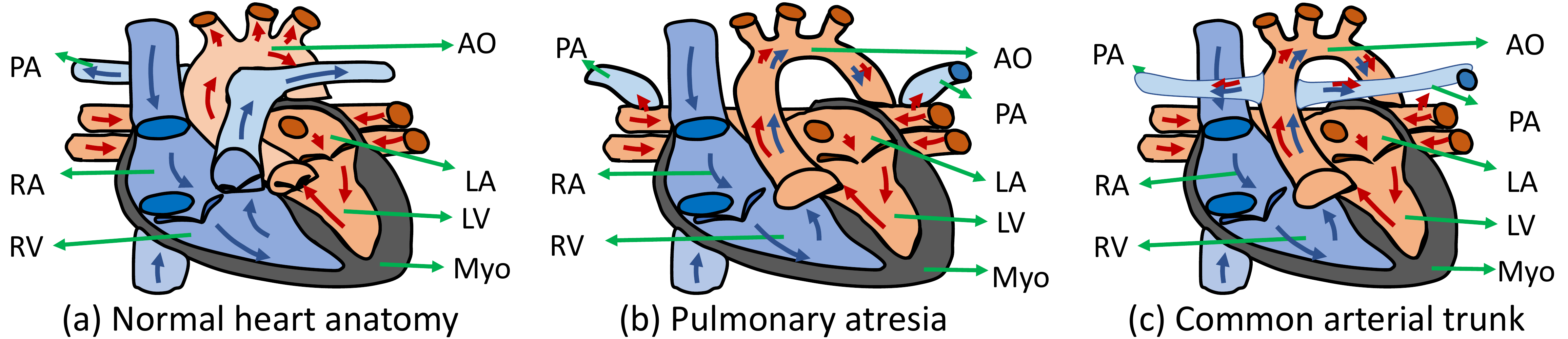}
\caption{Examples of large heart structure and great artery connection variations in CHD (LV-left ventricle, RV-right ventricle, LA-left atrium, RA-right atrium, Myo-myocardium, AO-aorta and PA-pulmonary artery). 
%In normal heart anatomy (a), the main trunk of PA is large. However, in pulmonary atresia (b) and common arterial trunk (c), the main trunk of PA disappears. 
Best viewed in color.
}
%\vspace{-32pt}
\label{challenge_example}
\end{figure}

On the other hand, automatic segmentation and 
classification of medical images in CHD is rather challenging.
%In addition to the common low contrast, noise, and other artifacts caused by various imaging modalities and techniques, 
Patients with CHD typically suffer from severe variation in heart structures and connections between different parts of the anatomy. Two examples are shown in Fig. \ref{challenge_example}:
%Compared with the normal anatomy in (a),
the disappearance of the main trunk of pulmonary artery (PA) in (b)(c) introduces much difficulty in the correct segmentation of PA and AO.
%On the other hand, 
In addition, CHD does not necessarily cause local tissue changes, as in lesions. 
As such, hearts with CHD have 
similar local statistics as normal hearts but with 
global structural changes. 
Automatic algorithms to detect the disorders need to be able to 
capture such changes, which require excellent usage of the contextual information. 
CHD classification is further complicated by the fact that 
%a patient, and thus his/her CT image, may exhibit more than one type 
a patient's CT image may exhibit more than one type
of CHD, and the number of types is more than 20 \cite{bhat2016illustrated}. 

%XX need to review works on heart disease classification; need to review
%our MICCAI paper. 
Various works exist in segmentation and classification 
of heart with {\em normal anatomy, e.g.,} \cite{wang2018two,payer2017multi,xu2018cfun,piccini2012respiratory,zhuang2016multi,dou2019pnp,ye2019multi,zheng2019hfa,liu2019automatic,zhou2019cross,zhang2019fine,habijan2019whole,wang2019msu}, most of which are based on deep neural networks (DNNs) \cite{xu2018quantization,wang2019scnn}.
%For example, \cite{payer2017multi} combines 3D U-net \cite{cciccek20163d} for segmentation and a simple convolutional neural network for label position prediction.
Recently, researchers started to explore heart segmentation in CHD.
The works \cite{yu20163d,wolterink2016dilated,yang2017hybrid,yang2017class,pace2018iterative} adopt 
DNNs for blood pool and myocardium segmentation only.
%Danielle \textit{et al.} \cite{pace2018iterative} uses iterative segmentation for left ventricle (LV) and aorta segmentation in CHD, which requires user interaction to locate an initial seed.
The only automatic whole heart and great artery 
segmentation method in CHD \cite{xu2019whole} in the literature 
uses a deep learning and shape similarity analysis based method. 
A 3D CT dataset for CHD segmentation is also released there. 
In addition to segmentation, there are also some works about classification of adult heart diseases \cite{bernard2018deep} but not CHD. The automatic classification of CHD still 
remains a missing
piece in the literature due to the complexity of CHD and the lack of dataset.

In this paper, we present ImageCHD, the first medical image dataset for CHD classification.
%attempt to tackle the CHD classification problem by combining segmentation and shape similarity analysis based on 3D CT images.
ImageCHD contains 110 3D Computed Tomography (CT) images which covers 16 types of CHD.
CT images are labelled by a team of four experienced cardiovascular radiologists with 7-substructure segmentation and CHD type classification.
The dataset is of decent size compared with other medical imaging datasets \cite{yu20163d}\cite{xu2018cfun}. 
%Considering the severe variations in structures and connections, ImageCHD represents a very hard problem that cannot be easily solved by any existing deep learning based computer vision frameworks today.
%we perform two subtasks: connection analysis among chambers and great arteries, and shape analysis of great arteries.
%(A 3D image in CHD contains one or more CHD features).
We also present a baseline method for automatic CHD classification based 
on the state-of-the-art CHD segmentation framework \cite{xu2019whole}, which is 
the first automatic CHD classification method in the literature.  
%Particularly, we first adopt \cite{xu2019whole} to segment the image, 
%based on which connection analysis is performed.
%Then, we obtain all the vessels from the segmentation results, and extract skeletons to match the templates in a library. The extracted connection and shape features from the two subtasks jointly decide the classification. 
Results show that the baseline framework can achieve a classification accuracy of 82.0\% under selective prediction scheme with 88.4\% coverage, and there is still big room for further improvement.

\begin{figure*}[!htb]
\centering
\includegraphics[width=0.95\textwidth]{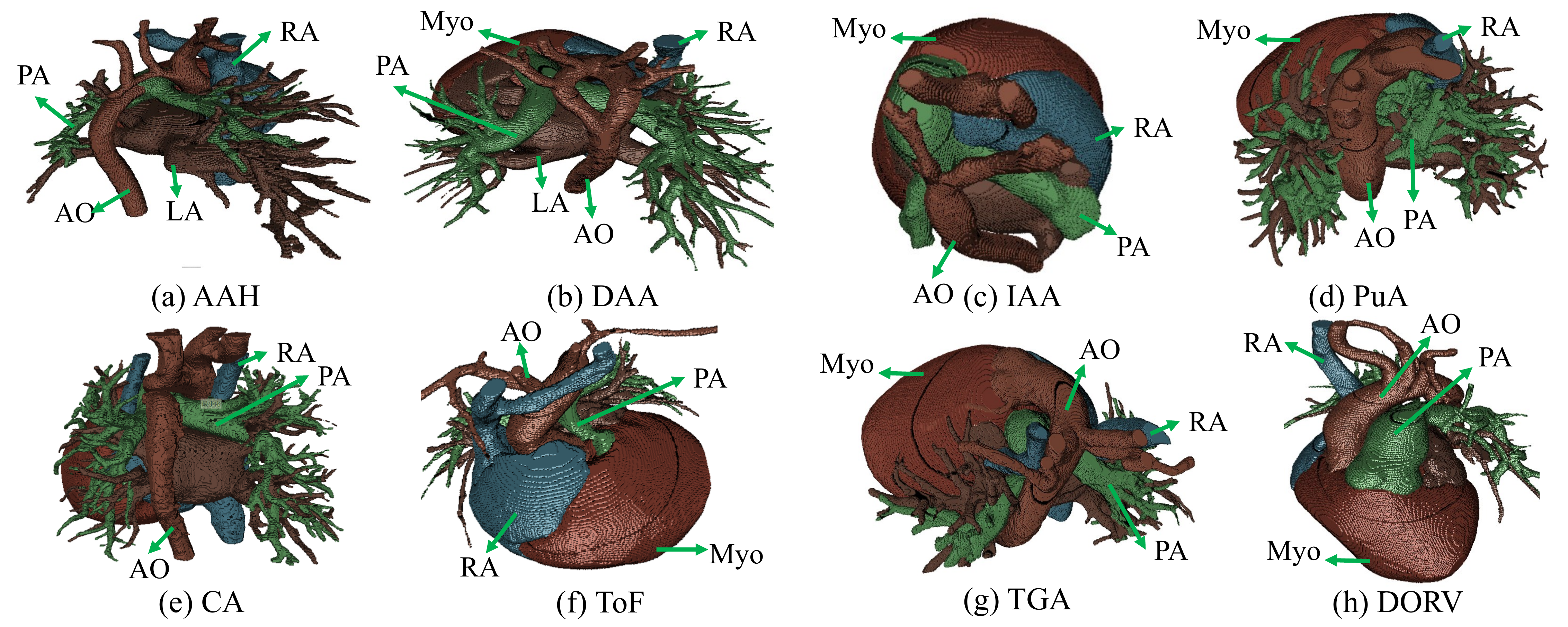}
%\vspace{-22pt}
\caption{Examples of CT images in the ImageCHD dataset with its types of CHD.
}
%\vspace{-12pt}
\label{fig_dataset}
\end{figure*}

\section{The ImageCHD Dataset}
The ImageCHD dataset consists of 3D CT images captured by a Siemens biograph 64 machine from 110 patients, with age between 1 month and 40 years (mostly between 1 month and 2 years). The size of the images is 512$\times$ 512$\times$(129-357), and the typical voxel size is 0.25$\times$0.25$\times$0.5$mm^3$. The dataset covers
16 types of CHD, which include eight common types (atrial septal defect (ASD), atrio-ventricular septal defect (AVSD), patent ductus arteriosus (PDA), pulmonary atresia (PuA), ventricular septal defect (VSD), co-arctation (CA), tetrology of fallot (TOF), and transposition of great arteries (TGA))
%shown in Table \ref{tab_commonCHD}
plus eight less common ones (pulmonary artery sling (PAS), double outlet right ventricle (DORV), common arterial trunk (CAT), double aortic arch (DAA), anomalous pulmonary venous drainage (APVC), aortic arch hypoplasia (AAH), interrupted aortic arch (IAA), double superior vena cava (DSVC)). The number of images associated with each is summarized in Table \ref{challenge_example}. 
%shoule be "table_dataset", interesting problem... challenge_example
Several examples of images in the dataset are shown in Figure \ref{fig_dataset}.
Due to the structure complexities, the labeling including segmentation and classification is performed by a team of four cardiovascular radiologists who have extensive experience with CHD. The segmentation label of each image is fulfilled by only one radiologist, and its diagnosis is performed by four.
%Each radiologist first independently diagnoses all the images. Then for each image, unless the four diagnoses all agree, consensus is reached through discussion. 
The time to label each image is around 1-1.5 hours on average.
%, 
%though some complicated ones can take much longer.
The segmentation include seven substructures: LV, RV, LA, RA, Myo, AO and PA.
%For easy processing, venae cavae and pulmonary vein are labeled as part of RA and LA respectively, as they are connected and their boundaries are relatively hard to define. %Anomalous vessels are labeled as one of the above seven substructures based on their connections.

\begin{table}%[]
\label{table_dataset}
\centering
{\small{
\caption{The types of CHD in the ImageCHD dataset (containing 110 3D CT images) and the associated number of images. Note that some images may correspond to more than one type of CHD.}
\begin{tabular}{ccccccccc}
\multicolumn{8}{c}{Common CHD} \\\hline
\centering
ASD&AVSD&VSD&TOF&PDA&TGA&CA&PuA  \\
32 & 18 & 44 & 12 & 14 & 5 & 6 & 16 \\
\hline
\multicolumn{8}{c}{Less Common CHD}&Normal\\\hline
\centering
PAS&DORV&CAT&DAA&APVC &AAH&IAA&DSVC \\
3 & 8 & 4 & 5         &6  &3 &3 &8 &6 \\\hline

\end{tabular}
% \begin{tabular}{p{0.55cm}p{0.75cm}p{0.55cm}p{0.55cm}p{0.55cm}p{0.55cm}p{0.55cm}p{0.55cm}p{0.75cm}}
% \hline
% \multicolumn{8}{c}{Common CHD}& \\\hline
% \centering
% ASD&AVSD&VSD&TOF&PDA&TGA&CA&PuA  \\
% 26&18&44&12&14&7&6&16 \\
% \hline
% \multicolumn{8}{c}{Less Common CHD}&Normal\\\hline
% \centering
% PAS&DORV&CAT&DAA&APVC &AAH&IAA&DSVC\\
% 3&8&4&5         &6  &3 &3&8&6 \\\hline
% \end{tabular}
%\vspace{-12pt}
}}
\end{table}

% \cite{xu2019whole} is the only fully automatic whole heart and great artery segmentation method for medical images in CHD in the literature. It demonstrates that 
% state-of-the-art deep learning based segmentation methods such as 
% Seg-CNN \cite{payer2017multi} for normal heart anatomy will fail in CHD. The method 
% instead uses deep learning to segment the four chambers and myocardium followed by blood pool, where variations are usually small, 
% and then extracts the connection information and apply graph matching to determine the categories of all the vessels. Based on a 
% CHD segmentation dataset with 
% 68 3D CT images covering 14 types of CHDs, the method 
% can increase Dice score by 12\% on average compared with the state-of-the-art whole heart and great artery segmentation method in normal anatomy. 

\section{The Baseline Method}

\begin{figure*}[htb]
\centering
\includegraphics[width=0.95\textwidth]{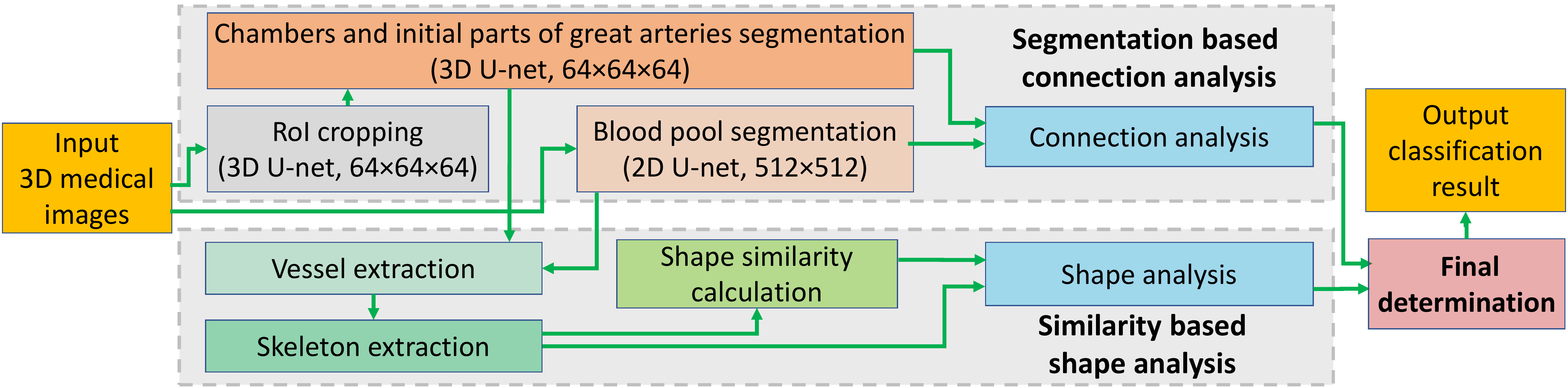}
%\vspace{-22pt}
\caption{Overview of the baseline method for CHD classification.
}
%\vspace{-12pt}
\label{fig_framework}
\end{figure*}

%\subsection{Overview}
\noindent\textbf{Overview:} Due to the lack of baseline method for CHD classification, 
along with the dataset we establish one 
 as shown in 
Fig. \ref{fig_framework}, which 
modifies and extends 
the whole heart and great artery segmentation method in CHD \cite{xu2019whole}. It
includes two subtasks: segmentation based connection analysis and similarity based shape analysis. Accordingly, the parts 
and connections most critical to the classification are extracted. 
%An example of the flow can be found in the supplementary material.

%% task/module

\begin{figure}[!tb]
\centering
\includegraphics[width=0.95\textwidth]{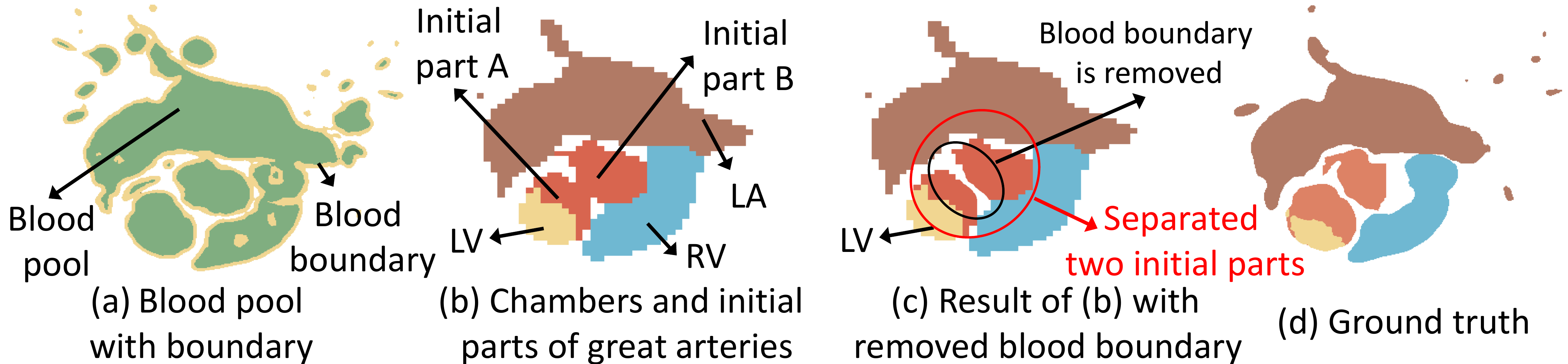}
%\vspace{-22pt}
\caption{Connection analysis between LV/RV and great arteries (AO and PA). %Best viewed in color.
}
%\vspace{-12pt}
\label{fig_connection12}
\end{figure}

\noindent\textbf{Segmentation based connection analysis}: Segmentation is performed with multiple U-Nets \cite{ronneberger2015u}.
There are two steps in segmentation: blood pool segmentation, and chambers and
initial parts of great arteries segmentation.
The former is fulfilled by a high-resolution (input size $512\times512$ ) 2D U-net, while the latter is performed with a 3D low-resolution (input size $64\times64\times64$ ) 3D U-net.
A Region of Interest (RoI) cropping is also included with a 3D U-net before the 3D segmentation.
%chambers and initial parts of great arteries segmentation.
With the segmentation results, connection analysis can be processed, which mainly extracts the connection features between great arteries (AO and PA) and LV/RV, and between LV/LA and RV/RA.
With the segmentation results, two connection analyses between chambers, AO and PA are then performed by the connection analysis module.
The first one analyzes the connections between LV/RV and great arteries. %xxxx
We remove high resolution boundary from low resolution substructures as shown in Fig. \ref{fig_connection12}(a)-(c). Compared with the ground truth in Fig. \ref{fig_connection12}(d), Fig. \ref{fig_connection12}(c) shows that the two initial parts are correctly separated (but not in (b) where they will be treated as connected).
The second one has a similar process as the first one.

\begin{figure}%[!htb]
\centering
\includegraphics[width=0.99\textwidth]{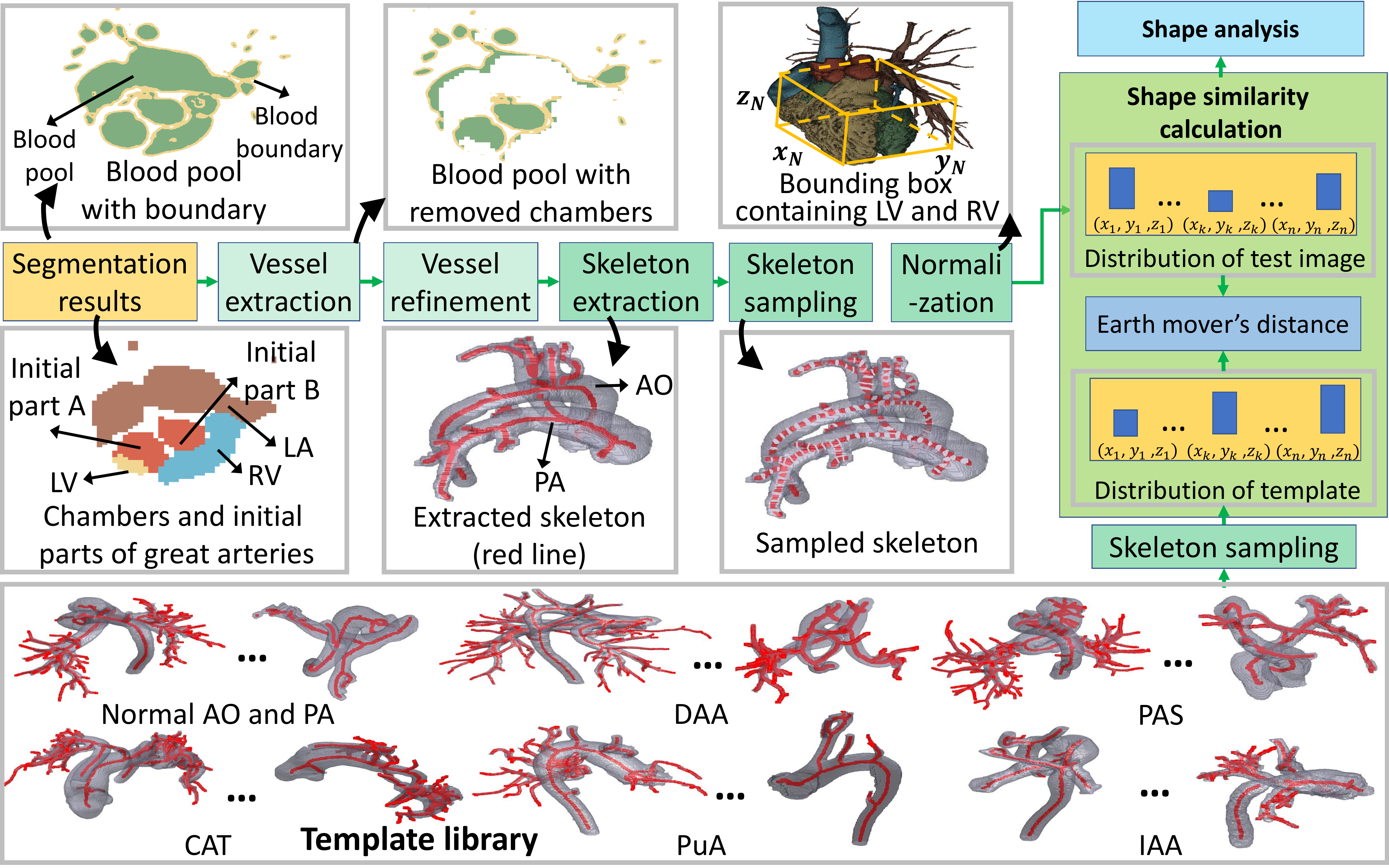}
%{figs/shape-flow.pdf}
%\vspace{-22pt}
\caption{Similarity based shape analysis of great arteries. Best viewed in color.
}
%\vspace{-12pt}
\label{fig_shape_flow}
\end{figure}

\noindent\textbf{Similarity based shape analysis}: 
The flow of this subtask is shown in Fig. \ref{fig_shape_flow}. With the segmentation results, vessel extraction removes the blood pool corresponding to chambers, and vessel refinement removes any remaining small islands in the image, and smooths it with erosion.
Then, the skeleton of the vessels are extracted, sampled, normalized, and fed to the shape similarity calculation module to obtain its similarity with all the templates in a pre-defined library.
Similarity module is performed using earth mover's distance (EMD) which is a widely used similarity metric for distributions \cite{rubner2000earth}.
Two factors need to be modeled: the $weight$ of each bin in the distribution, and the $distance$ between bins.
We model each sampled point in the sampled skeleton as a bin,
the Euclidean distance between the points as the distance between bins, and the volume of blood pool around the sampled point as the weight of its corresponding bin.
Particularly, the weight is defined as $r^3$ where $r$ is the radius of the inscribed sphere in the blood pool centered at the sampled point.
%and outputs the one with the highest similarity.
%Other shape features, such as whether a circle exists in the skeleton, are also extracted by analyzing the skeleton as a graph. These features are fed
The template library is manually created in advance 
and contains six categories of templates corresponding to five types of CHDs and the normal anatomy as shown in Fig. \ref{fig_shape_flow}, covering all the possible shapes of great arteries in our dataset. Each category contains multiple templates. 
%With the skeleton and its similarities with the templates, the shape analysis module outputs two kinds of features: the type of the template with the highest similarity, and selected skeleton features (e.g., whether a circle exists in the skeleton) useful for the classification of certain CHD types.
Finally, the shape analysis module takes the skeleton and its similarities to obtain two kinds of features.
The type of the template with the highest similarity is extracted as the first kind.
The second kind includes two skeleton features: whether a circle exists in the skeleton, and how the $r$ of the sample points varies.
These two features are desired because if there is a circle in the skeleton, the test image is with high possibility to be classified as DAA;
If a sampled point with a small $r$ is connected to two sampled points with a much larger $r$, narrow vessel happens, which is a possible indication of CA and PuA.

\noindent\textbf{Final determination}: With the extracted connection and shape features, the classification can be finally determined using a rule-based automatic approach. 
Specifically, ASD and VSD have unexpected connection between LA and RA, and LV and RV, respectively. AVSD is a combination of ASD and VSD, and the three can be classified according to the connection features between LA/LV and RA/RV.
DORV has two initial parts of great arteries, both of which are connected to RV.
TOF has connected LV and RV, as well as connected LV, RV and the initial part of AO. 
CHD with specific shapes including CAT, DAA, PuA, PAS and IAA as shown in Fig. \ref{fig_shape_flow} can be classified by their shape features.
PDA and CA are determined by analyzing the shapes and skeletons such as the variety of $r$ along the skeleton.
DSVC can be easily classified by analyzing the skeleton of RV, and APVC is determined by the number of islands that the LA has.
Note that if the connection and shape features do not fit any of the above rules, the classifier outputs \textit{uncertain} indicating that the test image cannot be handled and manual classification is needed.

\section{Experiment}

%\subsection{Experiment Setup}
\textbf{Experiment setup}: All the experiments run on a Nvidia GTX 1080Ti GPU with 11 GB memory.
We implement the 3D U-net using PyTorch based on \cite{payer2017multi}.
For 2D U-net, most configurations remain the same with those of the 3D U-net except that it adopts 5 levels and the number of filters in the initial level is 16.
Both Dice loss and cross entropy loss are used, and the training epochs are 2 and 480 for 2D U-net and 3D U-net, respectively.
Data augmentation and normalization are also adopted with the same configuration as in \cite{payer2017multi} for 3D U-net.
%Data normalization is the same as \cite{payer2017multi}.
%The learning rate is 0.0002 for the first 50\% epochs, and then 0.00002 afterward.
%Batch sizes of 4 and 8 are adopted for 3D and 2D U-net, respectively.
For both networks and all the analyses, three-fold cross validation is performed (about 37 images for testing, and 73 images for training). We split the dataset such that all types of CHD are present in each subset.
The classification considers a total of 17 classes, including 16 types of CHD and the normal anatomy.
The templates in the template library are randomly selected from the annotated training set.
%, though they may be not of the same type of CHD. The Dice score is used for segmentation evaluation.

In the evaluation, we use selective prediction scheme \cite{pidan2011selective} and report a case as uncertain if at least one chamber is missing (which does not correspond to any type in our dataset) in the segmentation results, or in the similarity calculation the minimum EMD is larger than 0.01.
%%xxxx Do we need to explain why this lead to uncertain? will be added in the journal version.
For these cases, manual classification by radiologists is needed.
%To show that it is indeed necessary to establish the baseline method 
%using the segmentation method in \cite{xu2019whole}, 
%we also create a modified baseline by replacing 
%it with a state-of-the-art segmentation method for heart in normal anatomy, Seg-CNN \cite{payer2017multi}, which 
%as established in \cite{xu2019whole} has lower Dice on CHD images. The same final determination module is used. 
%We train Seg-CNN with the same setup, and the input image is downsampled to 64$\times$64$\times$64.
To further evaluate how the baseline method performs against human experts, we also extract manual CT classification from the electronic health records (the manual results can still be wrong).

%invite five experienced cardiovascular radiologists to perform manual classification. Each radiologist independently classifies all the 110 images, and majority vote is used without any discussion (so the result can still be wrong).

%\subsection{Results and Analysis}
\noindent\textbf{Results and analysis}: The CHD classification result is shown in Table \ref{table_classification_result}.
%where each row corresponds to one ground truth class (16 types of CHD plus normal) and each column corresponds to one classified class (all ground truth classes plus uncertain).
Each entry (X, Y) in the table corresponds to the number of cases with ground truth class suggested by its row header and predicted class by its column header, where X, and Y are the results from the baseline, and those from radiologists respectively. Again, an image can contribute to multiple cases if it contains more than one types of CHD. From the table we can see that for the baseline method, due to segmentation error or feature extraction failure, 21 cases are classified as uncertain, yielding a 88.8\% coverage;
Out of the remaining 166 cases, 136 are correct.
Thus, for the baseline the overall classification accuracy is 72.7\% for full prediction, and 81.9\% for selective prediction.
For the modified baseline, the overall classification accuracy is 39.0\% for full prediction and 50.2\% for selective prediction.
%  u=42,c=74
On the other hand, the manual classification from experienced radiologist can achieve an overall accuracy of 90.5\%. It is interesting to note that out of the 17 classes, the baseline method achieves higher accuracy in one (PuA) and breaks even in five (VSD, TGA, CAT, DAA, and AAH) compared with manual classification.
In addition, Out of the 110 cases, the five radiologists only unanimously agreed on 78 cases, which further reflects the difficulty of the problem and the value of an automated tool.

The mean and standard deviation of Dice score of our baseline method for six substructures of chambers and initial parts of great vessels segmentation, and blood pool segmentation are shown in Table \ref{tab_result_substructure}.
We can notice that blood pool has the highest score, and initial parts of great vessels has the lowest, and the overall segmentation performance is moderate.
Though the segmentation performance of initial parts is low, its related types of CHDs (e.g., ToF, TGA) still achieve high classification accuracy which is due to the fact that only the critical segmentation determines the types of CHDs.
Comparing the performance of segmentation and classification, we can also notice that accurate segmentation usually helps classification, but not necessarily.

% \textbf{Comparison with the deep learning based method:} The baseline method can achieve about 32\% higher accuracy than the modified baseline in both the overall and the selective accuracy.
% This is due to the fact that the baseline method based on \cite{xu2019whole} 
% can well observe the structure variance while the modified one based on
% Seg-CNN cannot. An example is shown in Figure \ref{fig_compare_error}, 
% where the baseline can correctly classify the combined Ao and PA in CAT while the modified baseline cannot.
% This is because Seg-CNN only considers the surrounding pixels in 
% segmentation and does not consider the connection and shape features of the vessels.
% We can notice that the classification accuracy of vessel-caused CHDs (e.g., TGA, CAT) is much lower than that of chamber-caused CHDs (e.g., ASD, VSD, AVSD), which is due to the fact that the great arteries have more 
% significant variations than the chambers.

\noindent\textbf{Classification success:} Six types of CHD including TGA, CAT, DAA, AAH, PAS and PuA achieve relatively high accuracy, which is due to their clear and stable features that
distinguish them from normal anatomy. Such features can be easily captured
by either connection or shape features extracted by the baseline method.
For example, CAT has a main trunk that AO and PA are both connected to; DAA has a circular vessel which is composed of two aortic arches; PAS has a PA with very different shape; PuA has a very thin PA without main trunk;
AAH has a long period of narrow vessels in the arch; and TGA has a reversed connection to LV and RV.

\begin{table*}[tb]
{\scriptsize{
\centering
\caption{Number of cases (X, Y) with ground truth class and predicted class suggested by the row and column headers respectively, where X, and Y correspond to automatic classification by the baseline, and manual classification, respectively. Green numbers along the diagonal suggest correct cases. 
% Note that an image may contain, and also be classified as, more than one type so the numbers along each row may not add up to be the same for different frameworks due to wrong classification 
(U-Uncertain, 1-ASD, 2-AVSD, 3-VSD, 4-TOF, 5-PDA, 6-TGA, 7-CA, 8-IAA, 9-PAS, 10-DORV, 11-CAT, 12-DAA, 13-APVC,  14-AAH, 15-PuA, 16-DSVC, N-Normal)} 
\begin{tabular}{|p{0.65cm}||p{0.45cm}|p{0.69cm}|p{0.65cm}|p{0.68cm}|p{0.55cm}|p{0.55cm}|p{0.45cm}|p{0.45cm}|p{0.45cm}|p{0.45cm}|p{0.45cm}|p{0.45cm}|p{0.45cm}|p{0.45cm}|p{0.45cm}|p{0.65cm}|p{0.45cm}|p{0.55cm}|}
\hline
%\multicolumn{8}{c}{Common CHD}&Normal \\\hline
Type&U&1&2&3&4&5&6&7&8&9&10&11&12&13&14&15&16&N   \\\hline\hline
1&\textcolor{red}{6,0}   &\textcolor{ForestGreen}{18,24}&&&& &&&&&& &&&&& &\textcolor{red}{2,2} \\\hline
2&\textcolor{red}{3,0} &\textcolor{red}{1,3}&\textcolor{ForestGreen}{9,14}&\textcolor{red}{5,1}&& &&&&&& &&&&& & \\\hline
3&\textcolor{red}{1,0} &\textcolor{red}{1,2}&&\textcolor{ForestGreen}{42,42}&& &&&&&& &&&&& & \\\hline
4&\textcolor{red}{1,0} &&&\textcolor{red}{4,2}&\textcolor{ForestGreen}{7,10}& &&&&&& &&&&& & \\\hline
5& &&&&&\textcolor{ForestGreen}{7,14} &&&&&& &&&&& &\textcolor{red}{7,0} \\\hline

6& &&&&& &\textcolor{ForestGreen}{5,5}&&&&& &&&&& & \\\hline
7&\textcolor{red}{1,0} &&&&& &&\textcolor{ForestGreen}{4,6}&&&& &&&&\textcolor{red}{1,0}& & 
\\\hline
8& &&&&& &&&\textcolor{ForestGreen}{2,3}&& &&&&& & &\textcolor{red}{1,0} \\\hline
9&\textcolor{red}{1,0} &&&&& &&&&\textcolor{ForestGreen}{2,3}&& &&&&& & \\\hline
10&\textcolor{red}{1,0} &&&\textcolor{red}{3,1}&\textcolor{red}{1,1}& &&&&&\textcolor{ForestGreen}{3,6}& &&&&& & \\\hline

11& &&&&& &&&&&&\textcolor{ForestGreen}{4,4} &&&&& & \\\hline
12&  &&&&& &&&&&& &\textcolor{ForestGreen}{5,5}&&&& & \\\hline
13&\textcolor{red}{1,0} &&&&& &&&&&& &&\textcolor{ForestGreen}{3,6}&&& &\textcolor{red}{2,0} \\\hline
14&\textcolor{red}{1,0} &&&&& &&&&&& &&&\textcolor{ForestGreen}{2,2}&& &\textcolor{red}{0,1} \\\hline
15&\textcolor{red}{2,0} &&&&\textcolor{red}{0,2}& &\textcolor{red}{0,1}&&&&& &\textcolor{red}{0,1}&&&\textcolor{ForestGreen}{14,10}& & \\\hline

16&\textcolor{red}{1,0} &&&&& &&&&&& &&&&&\textcolor{ForestGreen}{5,7} &\textcolor{red}{2,1} \\\hline
N&\textcolor{red}{2,0} &&&&& &&&&&& &&&&& &\textcolor{ForestGreen}{4,6} \\\hline
%\vspace{-2pt}
% \begin{tabular}{|p{0.55cm}|p{0.25cm}|p{0.45cm}|p{0.25cm}|p{0.25cm}|p{0.25cm}|p{0.25cm}|p{0.25cm}|p{0.28cm}|p{0.28cm}|p{0.50cm}|p{0.25cm}|p{0.25cm}|p{0.5cm}|p{0.32cm}|p{0.25cm}|p{0.50cm}|p{0.60cm}|}
% \hline
% %\multicolumn{8}{c}{Common CHD}&Normal \\\hline
% &ASD&AVSD&VSD&TOF&PDA&TGA&CA&IAA&PAS&DORV&CAT&DAA&APVC &AAH&PuA&DSVC&Normal   \\\hline
% ASD   &17&&1&& &&&&&& &&&&& &2 \\\hline
% AVSD &1&9&5&& &&&&&& &&&&& & \\\hline
% VSD &1&&42&& &&&&&& &&&&& & \\\hline
% TOF &&&4&7& &&&&&& &&&&& & \\\hline
% PDA &&&&&7 &&&&&& &&&&& &7 \\\hline

% TGA &&&&& &6&&&&& &&&&& & \\\hline
% CA &&&&& &&4&&&& &&&&1& &
% \\\hline
% IAA &&&&& &&&2&& &&&&& & &1 \\\hline
% PAS &&&&& &&&&1&& &&&&& & \\\hline
% DORV &&&&& &&&&&3& &&&&& &4 \\\hline

% CAT &&&&& &&&&&&4 &&&&& & \\\hline
% DAA  &&&&& &&&&&& &5&&&& & \\\hline
% APVC &&&&& &&&&&& &&3&&& &2 \\\hline
% AAH &&&&& &&&&&& &&&2&& & \\\hline
% PuA &&&&& &&&&&& &&&&14& & \\\hline

% DSVC &&&&& &&&&&& &&&&&5 &2 \\\hline
% Normal &&&&& &&&&&& &&&&& & \\\hline
\end{tabular}
%\vspace{-12pt}
\label{table_classification_result}
}}
\end{table*}

\begin{table}[!tb]
\centering
%\vspace{-12pt}
\caption{Mean and standard deviation of Dice score of our baseline method (in \%) for six substructures of chambers and initial parts of great vessels segmentation, and blood pool segmentation.}
%\vspace{-5pt}
%\begin{tabular}{c{8cm}c{3cm}c{3cm}c{3cm}c{3cm}c{3cm}c{3cm}c{3cm}c{3cm}}
\begin{tabular}{ccccccccc}
%\multicolumn{8}{c}{Common CHD} & \multicolumn{6}{c}{Less Common CHD}&Normal\\
\hline
LV&RV&LA&RA&Initial parts of great vessels & Blood pool& Average  \\\hline
77.7&74.6&77.9&81.5&66.5&86.5&75.6\\
$\pm$16.2&$\pm$13.8&$\pm$11.2&$\pm$11.5&$\pm$15.1&$\pm$10.5&$\pm$10.2\\
\hline
\end{tabular}
\label{tab_result_substructure}
\end{table}

% The extracted skeleton in Fig.\ref{fig_uncertainty_error}(c) lacks left PA compared with the ground truth in (d). The left PA is removed during the vessel refinement because it is too thin.
% Such lack makes the normalization process allocate too heavy weights to the samples corresponding to right PA. As such, it has low similarity with any templates in the library.

% \begin{figure}[!tb]
% %%\vspace{-2pt}
% \includegraphics[width=0.45\textwidth]{figs/uncertainty-reason.pdf}
% %\vspace{-22pt}
% \caption{Examples of uncertain classification: low image contrast (a)(b) and feature extraction failure (c)(d). Best viewed in color.
% }
% \label{fig_uncertainty_error}
% \end{figure}

\begin{figure}[!tb]
\centering
\includegraphics[width=1\textwidth]{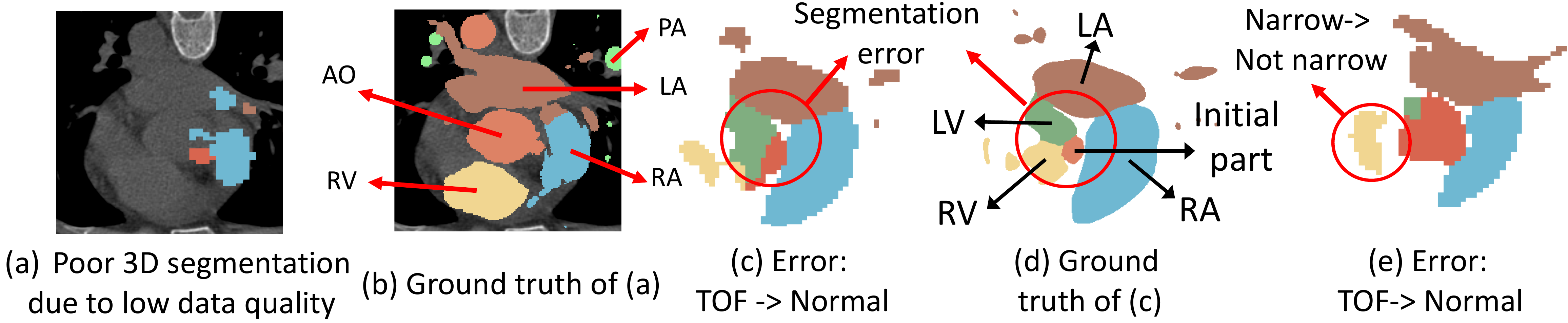}
%\vspace{-22pt}
\caption{Examples of classification failure: uncertain classification in (a-b), and wrong classification of TOF in (c) and (e). Best viewed in color.}
\label{fig_error}
\end{figure}

%%XX unify classification error and failure

\noindent\textbf{Classification failure:} Test images are classified as uncertain due to segmentation error.
Fig. \ref{fig_error} shows some examples of such error.
The test image in Fig.~\ref{fig_error}(a) has very low contrast, and its blood pool and boundary are not clear compared with other areas, resulting in segmentation error: compared with the ground truth in Fig. \ref{fig_error}(b), only RA and part of the initial parts of great arteries are segmented.
As for the cases where a CHD type is predicted but wrong, we will use TOF as examples, and leave the comprehensive discussion for all classes in the supplementary material.
Segmentation error around the initial parts of great arteries is the main reason of the classification failure of TOF as shown in Fig. \ref{fig_error}.
Compared with the ground truth in Fig. \ref{fig_error}(d), the 3D segmentation in Fig. \ref{fig_error}(c) labels part of LV as RV, resulting in the initial part only connected to RV rather than RV and LV.
As one of the main features of TOF is that one initial part is connected to both RV and LV, missing such feature leads to misclassification of TOF as VSD.
Another main feature of TOF is the narrow vessels in the initial part and its connected RV part, which can also lead to wrong classification if not detected correctly as shown in Fig. \ref{fig_error}(e). A precise threshold
to decide whether the vessels are narrow or not is still missing
even in clinical studies. 
%We hope that the baseline method can help to better establish such a threshold.

% Fig.\ref{fig_dorv_TOF_error}(d) shows an example of classification failure of DORV due to segmentation error.
% Compared with the ground truth in Fig.\ref{fig_dorv_TOF_error}(e), the thin vessel of RV is segmented as a combination of LV and RV, and the LV is connected to the initial part.
% Note that the small part of LV is also connected to the main part of LV which cannot be easily removed.
% Such segmentation error finally leads to the misclassification of DORV as normal.

\noindent\textbf{Discussion:}
We can notice that segmentation accuracy is important for successful classification of CHD. Higher segmentation accuracy can lead to better connection and shape feature extraction. 
%As connection and vessel extractions are most critical in the scenario of classification, they should be involved in the loss function design of the segmentation tasks. 
In addition, so far we have only considered the connection features in the blood pool and the shapes of the vessels. More structural 
features associated with classification should be considered to improve the performance, which 
due to the lack of local tissue changes, need innovations from 
the deep learning community and deeper collaboration between computer scientists and radiologists. 
%By releasing our dataset, we hope that future research can close the gap between the performance of the baseline method and that of experienced radiologists.

\section{Conclusion}
We introduce to the community the ImageCHD dataset \cite{ourdataset} in hopes of encouraging new research into unique, difficult and meaningful datasets. 
We also present a baseline method for comparison on this new dataset, 
based on a state-of-the-art whole-heart and great artery segmentation 
method for CHD images. 
Experimental results show that under selective prediction scheme the baseline method can achieve a classification accuracy of 81.9\%, 
leaving big room for improvement. 
We hope that the dataset and the baseline method can encourage 
new research that be used to better address not only the CHD 
classification but also a wider class of problems that have large global structural change but little 
local texture/feature change.

%the baseline method indicates that current deep learning cannot work well in ImageCHD, and combination of deep learning and traditional computer vision methods (such as our baseline method, or incorporating traditional methods in the loss function design in deep learning) may have the potential to solve this problem.
%To the best of the authors' knowledge, this is the first work on CHD classification. 
%To stimulate related research in the community, we have released the ImageCHD dataset to the public.
%The ImageNet \cite{deng2009imagenet} dataset has greatly pushed the boundaries of deep learning and computer vision, and we believe new areas such as diseEase diagnosis could be moved forward by unique, valuable, and challenging datasets. 

% \section{Acknowledgement}

% This work was approved by the Research Ethics Committee of Guangdong General Hospital, Guangdong Academy of Medical Science under Protocol No. 20140316.
% This work was supported by the Science and Technology Planning Project of Guangdong Province under Grant No. 2017A070701013, 2017B090904034, 2017B030314109, 2018B090944002, and 2019B020230003, Guangdong peak project under Grant No. DFJH201802, the National Key Research and Development Program under Grant No. 2018YFC1002600, the Natural Science Foundation of Guangdong Province under Grant No. 2018A030313785.

\bibliographystyle{splncs04}
\bibliography{reference}

\begin{thebibliography}{10}
\providecommand{\url}[1]{\texttt{#1}}
\providecommand{\urlprefix}{URL }
\providecommand{\doi}[1]{https://doi.org/#1}

\bibitem{ourdataset}
https://github.com/XiaoweiXu/ImageCHD-A-3D-Computed-Tomography-Image-Dataset-for-Classification-of-Congenital-Heart-Disease

\bibitem{bernard2018deep}
Bernard, O., Lalande, A., Zotti, C., Cervenansky, F., Yang, X., Heng, P.A.,
  Cetin, I., Lekadir, K., Camara, O., Ballester, M.A.G., et~al.: Deep learning
  techniques for automatic mri cardiac multi-structures segmentation and
  diagnosis: Is the problem solved? IEEE transactions on medical imaging
  \textbf{37}(11),  2514--2525 (2018)

\bibitem{bhat2016illustrated}
Bhat, V., BeLaVaL, V., Gadabanahalli, K., Raj, V., Shah, S.: Illustrated
  imaging essay on congenital heart diseases: multimodality approach part i:
  clinical perspective, anatomy and imaging techniques. Journal of clinical and
  diagnostic research: JCDR  \textbf{10}(5), ~TE01 (2016)

\bibitem{dou2019pnp}
Dou, Q., Ouyang, C., Chen, C., Chen, H., Glocker, B., Zhuang, X., Heng, P.A.:
  Pnp-adanet: Plug-and-play adversarial domain adaptation network at unpaired
  cross-modality cardiac segmentation. IEEE Access  \textbf{7},  99065--99076
  (2019)

\bibitem{habijan2019whole}
Habijan, M., Leventi{\'c}, H., Gali{\'c}, I., Babin, D.: Whole heart
  segmentation from ct images using 3d u-net architecture. In: 2019
  International Conference on Systems, Signals and Image Processing (IWSSIP).
  pp. 121--126. IEEE (2019)

\bibitem{liu2019automatic}
Liu, T., Tian, Y., Zhao, S., Huang, X., Wang, Q.: Automatic whole heart
  segmentation using a two-stage u-net framework and an adaptive threshold
  window. IEEE Access  \textbf{7},  83628--83636 (2019)

\bibitem{pace2018iterative}
Pace, D.F., Dalca, A.V., Brosch, T., Geva, T., Powell, A.J., Weese, J.,
  Moghari, M.H., Golland, P.: Iterative segmentation from limited training
  data: Applications to congenital heart disease. In: Deep Learning in Medical
  Image Analysis and Multimodal Learning for Clinical Decision Support, pp.
  334--342. Springer (2018)

\bibitem{payer2017multi}
Payer, C., {\v{S}}tern, D., Bischof, H., Urschler, M.: Multi-label whole heart
  segmentation using cnns and anatomical label configurations. In:
  International Workshop on Statistical Atlases and Computational Models of the
  Heart. pp. 190--198. Springer (2017)

\bibitem{piccini2012respiratory}
Piccini, D., Littmann, A., Nielles-Vallespin, S., Zenge, M.O.: Respiratory
  self-navigation for whole-heart bright-blood coronary mri: methods for robust
  isolation and automatic segmentation of the blood pool. Magnetic resonance in
  medicine  \textbf{68}(2),  571--579 (2012)

\bibitem{pidan2011selective}
Pidan, D., El-Yaniv, R.: Selective prediction of financial trends with hidden
  markov models. In: Advances in Neural Information Processing Systems. pp.
  855--863 (2011)

\bibitem{ronneberger2015u}
Ronneberger, O., Fischer, P., Brox, T.: U-net: Convolutional networks for
  biomedical image segmentation. In: International Conference on Medical image
  computing and computer-assisted intervention. pp. 234--241. Springer (2015)

\bibitem{rubner2000earth}
Rubner, Y., Tomasi, C., Guibas, L.J.: The earth mover's distance as a metric
  for image retrieval. International journal of computer vision
  \textbf{40}(2),  99--121 (2000)

\bibitem{wang2018two}
Wang, C., MacGillivray, T., Macnaught, G., Yang, G., Newby, D.: A two-stage 3d
  unet framework for multi-class segmentation on full resolution image. arXiv
  preprint arXiv:1804.04341  (2018)

\bibitem{wang2019msu}
Wang, T., Xiong, J., Xu, X., Jiang, M., Yuan, H., Huang, M., Zhuang, J., Shi,
  Y.: Msu-net: Multiscale statistical u-net for real-time 3d cardiac mri video
  segmentation. In: International Conference on Medical Image Computing and
  Computer-Assisted Intervention. pp. 614--622. Springer (2019)

\bibitem{wang2019scnn}
Wang, T., Xiong, J., Xu, X., Shi, Y.: {SCNN}: A general distribution based
  statistical convolutional neural network with application to video object
  detection. arXiv preprint arXiv:1903.07663  (2019)

\bibitem{wolterink2016dilated}
Wolterink, J.M., Leiner, T., Viergever, M.A., I{\v{s}}gum, I.: Dilated
  convolutional neural networks for cardiovascular mr segmentation in
  congenital heart disease. In: Reconstruction, segmentation, and analysis of
  medical images, pp. 95--102. Springer (2016)

\bibitem{xu2018quantization}
Xu, X., Lu, Q., Yang, L., Hu, S., Chen, D., Hu, Y., Shi, Y.: Quantization of
  fully convolutional networks for accurate biomedical image segmentation. In:
  Proceedings of the IEEE conference on computer vision and pattern
  recognition. pp. 8300--8308 (2018)

\bibitem{xu2019whole}
Xu, X., Wang, T., Shi, Y., Yuan, H., Jia, Q., Huang, M., Zhuang, J.: Whole
  heart and great vessel segmentation in congenital heart disease using deep
  neural networks and graph matching. In: International Conference on Medical
  Image Computing and Computer-Assisted Intervention. pp. 477--485. Springer
  (2019)

\bibitem{xu2018cfun}
Xu, Z., Wu, Z., Feng, J.: Cfun: Combining faster r-cnn and u-net network for
  efficient whole heart segmentation. arXiv preprint arXiv:1812.04914  (2018)

\bibitem{yang2017class}
Yang, X., Bian, C., Yu, L., Ni, D., Heng, P.A.: Class-balanced deep neural
  network for automatic ventricular structure segmentation. In: International
  Workshop on Statistical Atlases and Computational Models of the Heart. pp.
  152--160. Springer (2017)

\bibitem{yang2017hybrid}
Yang, X., Bian, C., Yu, L., Ni, D., Heng, P.A.: Hybrid loss guided
  convolutional networks for whole heart parsing. In: International Workshop on
  Statistical Atlases and Computational Models of the Heart. pp. 215--223.
  Springer (2017)

\bibitem{ye2019multi}
Ye, C., Wang, W., Zhang, S., Wang, K.: Multi-depth fusion network for
  whole-heart ct image segmentation. IEEE Access  \textbf{7},  23421--23429
  (2019)

\bibitem{yu20163d}
Yu, L., Yang, X., Qin, J., Heng, P.A.: 3d fractalnet: dense volumetric
  segmentation for cardiovascular mri volumes. In: Reconstruction,
  segmentation, and analysis of medical images, pp. 103--110. Springer (2016)

\bibitem{zhang2019fine}
Zhang, R., Chung, A.C.: A fine-grain error map prediction and segmentation
  quality assessment framework for whole-heart segmentation. In: International
  Conference on Medical Image Computing and Computer-Assisted Intervention. pp.
  550--558. Springer (2019)

\bibitem{zheng2019hfa}
Zheng, H., Yang, L., Han, J., Zhang, Y., Liang, P., Zhao, Z., Wang, C., Chen,
  D.Z.: Hfa-net: 3d cardiovascular image segmentation with asymmetrical pooling
  and content-aware fusion. In: International Conference on Medical Image
  Computing and Computer-Assisted Intervention. pp. 759--767. Springer (2019)

\bibitem{zhou2019cross}
Zhou, Z., Guo, X., Yang, W., Shi, Y., Zhou, L., Wang, L., Yang, M.: Cross-modal
  attention-guided convolutional network for multi-modal cardiac segmentation.
  In: International Workshop on Machine Learning in Medical Imaging. pp.
  601--610. Springer (2019)

\bibitem{zhuang2016multi}
Zhuang, X., Shen, J.: Multi-scale patch and multi-modality atlases for whole
  heart segmentation of mri. Medical image analysis  \textbf{31},  77--87
  (2016)

\end{thebibliography}

\end{document}